 \documentclass[aps,twocolumn,showpacs,floatfix,prl]{revtex4-1}
\usepackage{graphicx}
\usepackage{dcolumn}
\usepackage{bm}

\usepackage{amsmath,amssymb}

\usepackage[latin1]{inputenc}

\usepackage{color,soul}
\setstcolor{red}

\protect

\newcommand{\be}{\begin{equation}}
\newcommand{\ee}{\end{equation}}

\begin{document}

\title{Exploration and Trapping of Mortal Random Walkers}

\author{S. B. Yuste$^{1}$, E. Abad$^{2}$, and Katja Lindenberg$^{3}$}
\affiliation{$^{(1)}$
Departamento de F\'{\i}sica, Universidad de Extremadura,
E-06071 Badajoz, Spain\\
$^{(2)}$Departamento de F\'{\i}sica Aplicada, Centro Universitario de M\'erida,
Universidad de Extremadura, E-06800 M\'erida, Spain\\
$^{(3)}$Department of Chemistry and Biochemistry, and BioCircuits Institute,
University of California San Diego, 9500 Gilman Drive, La Jolla, CA
92093-0340, USA}

\date{\today}
\begin{abstract}
Exploration and trapping properties of random walkers that may evanesce at any time as they walk have seen very little treatment in the literature, and yet a finite lifetime is a frequent occurrence, and its effects on a number of random walk properties may be profound.  For instance, whereas the average number of distinct sites visited by an immortal walker grows with time without bound, that of a mortal walker may, depending on dimensionality and rate of evanescence, remain finite or keep growing with the passage of time. This number can in turn be used to calculate  other classic quantities such as the survival probability of a target surrounded by diffusing traps.  If the traps are immortal, the survival probability will vanish with increasing time.  However, if the traps are evanescent, the target may be spared a certain death.  We analytically calculate a number of basic and broadly used quantities for evanescent random walkers.
\end{abstract}


\maketitle

Random walk models provide a quintessential approach to transport and related processes in condensed media and
have been studied for more than a century - indeed, as an antecedent, there is reference to probability and statistical inference
in biblical texts~\cite{Bib2}.
It is therefore surprising to find important problems in this arena that have not yet been explored, especially ones that are broadly applicable and that can be dealt with analytically.  This letter deals with one class of such problems associated with the territory explored by {\em mortal} or {\em evanescent} random walkers.
Mortal walkers or mortal
diffusing particles may disappear in the course of their motion. This disappearance may, for instance, be the result of a finite walker lifetime such as in a unimolecular reaction or a natural decay process.  Other examples of disappearance events may arise from an encounter of a walker with another walker leading to the annihilation of one or both, as may occur in
radical recombination, in exciton trapping in photosynthesis, or in growth by aggregation. Many references to such  phenomena can be found in \cite{Rice1985,Sano1979,Kotomin1996}, and most recently in \cite{Eliazar2012a}.
On occasion but only rarely one can find models in the literature where evanescence is incorporated in an explicit way.
One interesting instance involves molecular motors that may detach irreversibly from the transport track \cite{Kolomeisky2000a}.
The underlying random walk model has been generalized
in
\cite{Kolomeisky2000b,Kolomeisky2009}.

The statistical properties of the territory explored by {\em immortal} random walkers as a function of time have been studied in a variety of contexts
\cite{Larralde1992a,Yuste1999,Acedo2000,Weiss2002,Majumdar2012}.
So have related quantities such as the probability of return to a given location.  In turn, there are further connections between these and reaction kinetic quantities such as the survival probability of a target particle surrounded by diffusing traps (``target problem")
\cite{Hughes1995,Benichou2000,Condamin2008,Benichou2010}.
Early last century these problems experienced
a surge in the literature with the pioneering work of Smoluchowski on
diffusion-limited chemical kinetics.
More recently, a resurgence of interest
started with the classic works of Scher et al. on stochastic transport in amorphous condensed media
\cite{Scher1973a,Scher1975}.
Traps or defects in these media are slowed down by the disordered environment and consequently experience so-called anomalous diffusion.  This slowing down, when incorporated in the ``defect diffusion model," leads to
stretched exponential relaxation, which turns out to be widely ubiquitous in nature
\cite{Scher1991,Shlesinger1988a,Eliazar2012a}.
The periodic surge of
interest has
again been proved by the plethora of recent books and chapters on anomalous diffusion models as a descriptive tool of crowded disordered condensed systems
\cite{AnotransBook2008,Sokolov2012}.

{\bf Statement of the problem.}
We consider a symmetric nearest-neighbor random walk on a $d$-dimensional lattice, that is, a P\'olya walk.  The walker steps at discrete times $t_n$, where $n$ is the number of steps.
We will also consider the continuous version,
a diffusive process in a continuous medium taking place in continuous time.
Our first goal is to calculate a quantity which can then be used to calculate many others: the average number $S_n^*$ of distinct sites visited by an evanescent walker up to time $t_n$.
The corresponding continuum quantity is the average volume $v_t^*$ of the Wiener sausage generated by mortal particles up to time $t$. The asterisks denote evanescent particles (the corresponding quantities for immortal walkers are indicated without an asterisk). In turn, these results can be used to address other classic problems, now for mortal walkers.  Perhaps one of the most interesting arises from the well-known connection between the survival probability up to a given step number or time of a target particle surrounded by a concentration of diffusive evanescent traps.  For immortal walkers, in the discrete problem this survival probability is $\phi_n = \exp[-\rho (S_n-1)]$ and in the continuous case it is $\phi(t)= \exp(-cR^dv_t)$, where $\rho$ and $c$ denote the density of walkers in appropriate units and $R$ is the radius of the target (assuming point traps; otherwise $R$ is the sum of the radii of the target and a trap). These relations persist for mortal walkers.

To arrive at the number of distinct sites visited by an evanescent walker,
we introduce the probability $P^*_{m,n-m}(s|s')$ of finding an evanescent walker at
site $s$ after
$n-m$ steps if the walk started at
$s'$ at step $m$.
The probability $P_{0,n}^*(s|s_0)$  is
the outcome of carrying out the experiment repeatedly, starting the walker at site $s_0$ and counting the fraction of realizations that arrive at site $s$ at step $n$. Alternately and equivalently, if a number of noninteracting walkers all start at step $0$ at site $s_0$, this is the fraction that arrive at site $s$ after $n$ steps.
The probability $P_{m,n}^*(s|s')$ is related to the corresponding well-studied probability $P_{m,n}(s|s')$ for immortal walkers as
$P_{m,n}^*(s|s') = [\rho(n)/\rho(m)] P_{m,n}(s|s')$,
where $\rho(n)$ is the fraction of realizations for which the walker has not evanesced up to step $n$ or, alternately, the concentration of walkers that have not evanesced up to that step (with $\rho(0)=1$). We also introduce $F_n^*(s|s_0)$, the probability that the evanescent walker arrives at site $s$ {\em for the first time} at step $n$ if the walker started at site $s_0$ at step $n=0$.
The probabilities $P_{m,n}^*$ and $F_n^*$ are related in the same way as for immortal walkers:
\be
P_{0,n}^*(s|s_0) = \delta_{ss_0}\delta_{n0} + \sum_{j=1}^n F_j^*(s|s_0)P_{j,n-j}^*(s|s), \quad n \geq 0.
\label{PFP}
\ee

Let $\Delta_n^*$ denote the average number of {\em new} sites
visited by the $n^{th}$ step of an evanescent walk, with $\Delta_0^*=1$. Then
\be
\label{SDelta}
S^*_n  =\sum_{j=0}^n  \Delta^*_j  =\sum_{j=0}^n\rho(j)\,   \Delta_j .
\ee
We define the generating function of any $n$-dependent quantity $A_n(\cdot)$ by
$A(\cdot;\xi) \equiv \sum_{n=0}^\infty A_n(\cdot)\xi^n$.
 The generating functions of $S_n^*$ and $\Delta_n^*$ are
related by
\be
\label{SD2}
S^*(\xi)= \frac{\Delta^*(\xi)}{1-\xi}.
\ee
On the other hand,
 $\Delta^*_n =\sum_{s\neq s_0} F^*_n(s|s_0),~~  n \neq 0$.
Multiplying by $\xi^n$, summing over $n$, and reversing the order of summation yields
\be
\label{DF}
\Delta^*(\xi) = 1+\sum_{s\neq s_0}   F^*(s|s_0;\xi).
\ee

In order to go further we need to specify particular forms of evanescence. We consider exponential and power-law decay of the concentration of walkers.  The former is the typical unimolecular decay that describes spontaneous death;  the latter is
associated with more complex chemical reactions
\cite{Kotomin1996,Chakravarty2009}.

{\bf Exponential evanescence.} With exponential evanescence,  $\rho(n) = \exp(-\lambda n)$.
For immortal walkers on a regular lattice,  the walk is time invariant, $P_{j,n}(s|s_0)=P_{n-j}(s|s_0)$.
Exponential evanescence is the only form
that preserves this property for $P^*$.  It then follows from Eq.~(\ref{PFP})
that
 \be
\label{FPxi}
F^*(s|s_0;\xi)= \frac{P^*(s|s_0;\xi)-\delta_{ss_0}}{P^*(0;\xi)},
\ee
where
translational invariance
implies that $P^*(s|s;\xi)=P^*(s_0|s_0;\xi)\equiv P^*(0;\xi)$.
Then, from Eq.~(\ref{FPxi})  with Eqs.~(\ref{SD2}) and (\ref{DF}) one finds
\be
\label{SD3}
S^*(\xi)= \frac{1}{1-\xi} \sum_{s} \frac{P^*(s|s_0;\xi)}{P^*(0;\xi)}
\ee
and,
using the abbreviated notation $\hat \xi= e^{-\lambda} \xi$,
$(1-\xi)\,S^*(\xi)=[(1-\hat \xi)P(0;\hat \xi)]^{-1}$.
Here we have used the relations $\sum_{s}P^*(s|s_0;\xi)=\rho(\xi)=1/(1-\hat \xi)$ and $P^*(0;\xi)=P(0;\hat\xi)$. Lattice Green functions $P(0;\xi)$ are well known for the most relevant $d$-dimensional lattices \cite{Hughes1995,Guttmann2010}, which then allows us to find a number of results for the evanescent walk.
The expansion of $S^*(\xi)$
in a power series yields
the average number of sites visited up to time $t_n$ by a mortal P\'olya walker with exponential evanescence. For immortal walkers $S_n\to\infty$ as $n\to\infty$. For mortal walkers  in the case of exponential evanescence it is \emph{finite}, and is given by
\be
\label{Sinfgral}
S_\infty^* =  \frac{1}{1-e^{-\lambda}} \;\frac{1}{P(0;e^{-\lambda})}.
\ee
Specific values of $S_\infty^*$ depend on dimension and type of lattice.
For $d=1$, $P(0;\xi)=(1-\xi^2)^{-1/2}$, so that
$S_\infty^*= \left[(1+e^{-\lambda})/(1-e^{-\lambda})\right]^{1/2}$.
For a two-dimensional square lattice $P(0;\xi)=2K(\xi)/\pi$, where $K(\cdot)$ is the elliptic integral of the first kind.

From the known asymptotic behaviors of $P(0;\xi)$ as $\xi\to 1^-$ and the fact that $\hat{\xi}=e^{-\lambda} \xi$, one can
arrive at the large-$n$ behavior of $S_n^*$ for
slow evanescence ($\lambda \to 0$). Focusing
on the leading asymptotic contribution, we note that
for $d=2$,
$P(0;\xi\to 1^-)\sim A/\pi \ln[B/(1-\xi)]$, where the constants $A$ and $B$ depend on the type of lattice \cite{Hughes1995}. This behavior in Eq.~(\ref{Sinfgral})
yields
$ S_\infty^* \sim   \pi/[\lambda\, A \log(B/\lambda)]$ as $\lambda\to 0$.
In dimension $d\ge 3$ the probability
that a walker returns to the origin is
$\mathcal{R}=1-1/P(0;1)$.
Hence,
$S_\infty^* \sim (1-\mathcal{R}) \lambda^{-1}$ as $\lambda\to 0$.

\begin{figure}
\begin{center}
\includegraphics[width=0.48\textwidth]{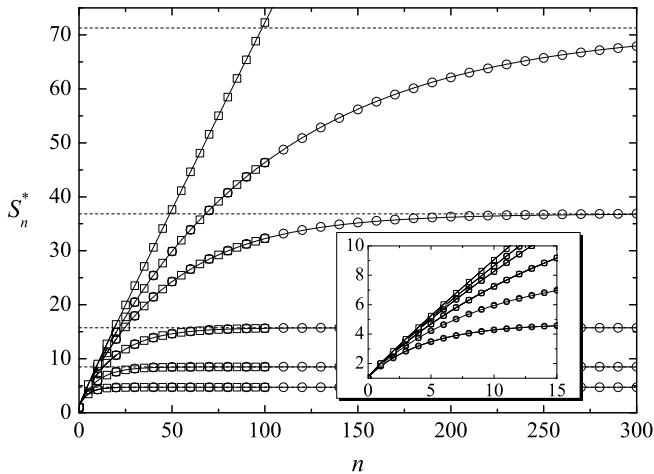}
\caption{
 $S_n^*$  vs \ $n$ for the simple cubic lattice and, from top to bottom, $\lambda=0, 0.01, 0.02, 0.05, 0.1, 0.2$. Solid lines:
Eq.~\eqref{Sn3d4}; broken lines: $S_\infty^*$ given by Eq.~\eqref{Sinfgral}; squares: exact values obtained by identifying the 100 first coefficients in the $\xi$-power expansion of $S^*(\xi)$; circles: simulation values for $10^5$ runs. The excellent performance of the asymptotic expression \eqref{Sn3d4} even for small $n$ is shown in the inset.  For the simple cubic lattice
$u_0=1/(1-\mathcal{R})\simeq 1.51639$
and $u_1=3^{3/2}/(\pi\sqrt{2})$.
\label{fig1}
}
\end{center}
\end{figure}

The approach of
$S_n^*$ to
$S_\infty^*$ for large $n$ follows
from the subdominant behavior of $P(0;\xi)$ as $\xi\to 1^-$.
For three-dimensional lattices
$P(0;\xi) =\sum_{m=0}^{\infty} (-1)^m u_m (1-\xi)^{m/2}=\left[\sum_{m=0}^{\infty}   v_m (1-\xi)^{m/2}\right]^{-1}$, where the $u_m$ and $v_m$
are known for a number of lattices \cite{Montroll1965,Hughes1995}.  Using the second expression in the result following Eq.~(\ref{SD3}), expanding in powers of $\xi$,
and retaining only the first two terms leads to
\begin{align}
\label{Sn3d4}
 S_n^*   &\sim S_\infty^*   -  \frac{1}{u_0}    \frac{ e^{-\lambda (n+1)}}{1-e^{-\lambda}}  -
\frac{u_1}{u_0^2}  \frac{ I_{e^{-\lambda}}(n+1,1/2)}{(1-e^{-\lambda})^{1/2}}
\end{align}
for $n \to \infty$ ,  where $I_{x}(a,b)$  is the regularized Beta function.  This asymptotic expression turns out to be surprisingly accurate even for relatively small  $n$ and for $\lambda$'s that need not be extremely small (Fig. \ref{fig1}).  In fact, the results for $\lambda$ close to zero are so good that
one can find the large-$n$ asymptotic expression for $S_n$ by taking the limit $\lambda \to 0$ of Eq.~\eqref{Sn3d4}.
Expanding this result in powers of $n$ yields a series whose first three terms
(proportional to $n$, $n^{1/2}$, and $n^0$) are identical to those obtained by expanding the exact result for $S_n$ \cite{Montroll1965}. Differences only appear in the fourth term, proportional to $n^{-1/2}$.

\begin{figure}
\begin{center}
\includegraphics[width=0.45\textwidth]{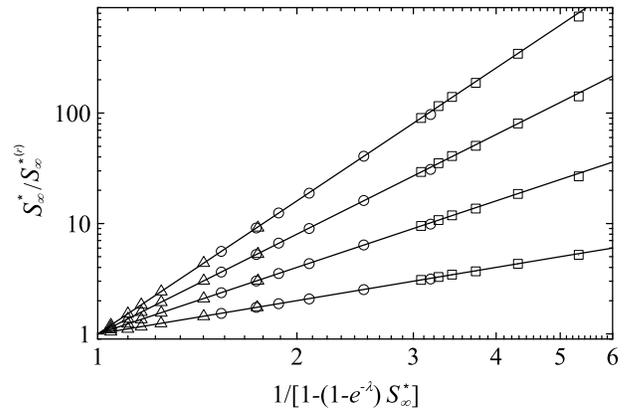}
\caption{\label{fig:SnrSn1}
$S_\infty^{*(r)}$ vs  $S_\infty^{*}$ and $\lambda$ for several values of $r$ and  $\lambda$ and three different lattices. Symbols: numerical simulations for $d=1$ (triangles), $d=2$ (square lattice, circles), and  $d=3$ (cubic lattice, squares) for $10^5$ runs. The values of $S_\infty^{*(r)}$ from the simulation of $S_n^{*(r)}$ with $n$ sufficiently large to observe no change in at least three significant figures. From left to right:  $\lambda=0.1, 0.05, 0.03, 0.01, 0.05, 0.001$, with  from top to bottom $r=2, 3, 4, 5$. The straight lines of slope $(r-1)$ through the origin are the theoretical predictions. }
\end{center}
\end{figure}

A quantity related to $S_n^*$ is $S_n^{*(r)}$, the average number of sites revisited at least $r$ times by an evanescent walker in an $n$-step walk.  Following the procedure in
\cite{Montroll1965},
one finds that  the generating function for this number for exponentially evanescent mortal walkers is given by
 \be
\label{Sr2}
 S^{*(r)}(\xi)=
  \left[1- \frac{1}{P(0; \hat{\xi)}}\right]^{r-1} S^{*}(\xi).
\ee
From here one  finds
$S_n^{*(r)}$ in terms of $S_n^{*}$ \cite{Montroll1965}. For instance, in dimension $d=1$,
$S_n^{*(2)} =  S_n^{*}-1-e^{-\lambda}$,
$S_n^{*(3)} = 2 S_n^{* }  -e^{-2\lambda}   S_n^{* }  - 2 -2e^{-\lambda}$, etc.
The average number of sites visited $r$ times before the walker dies is in any dimension given  by
$S_\infty^{*(r)}=[(1-e^{-\lambda}) \mu_\infty^*]^{r-1} \, (S_\infty^*)^r$, is shown
as a function of $r$ and of $\lambda$  in Fig.~\ref{fig:SnrSn1}
and compared extremely favorably with simulation results. For the average number of revisits to the origin after $n$ steps, $\mu_n^*$, one finds the generating function
$(1-\xi)\,\mu^*(\xi)=    P(0;\hat{\xi}) -1$.  (We follow the convention of \emph{not} counting the initial occupancy of the origin as the first revisitation \cite{Hughes1995}). We find that
in any dimension
$\mu_\infty^*=[(1-e^{-\lambda}) S_\infty^*]^{-1}-1$, and the average number of visits to a site $s$ other than the origin
is $\mu_\infty^*(s|s_0) =
P(s|s_0;e^{-\lambda})$, one of the few previously known results for exponentially evanescent walkers (see Sec. 3.2.4 of Ref.~\cite{Hughes1995}).

{\bf Power law evanescence.}
Power law evanescence
is given by $\rho(n)=(1+\lambda n)^{-\beta}$ with $\lambda>0$ and  $\beta>0$. Here it is convenient to  directly use the relation (\ref{SDelta}) and rely on the knowledge of  $\Delta_n$ for large and small $n$ for the most common lattices \cite{Hughes1995,Montroll1965}. For example, $\Delta_n\sim
  (1-\mathcal{R})\, \lambda^{-\beta} n^{-\beta} \left(1+C\, n^{-1/2} +\cdots\right)$ for three-dimensional lattices and large $n$.
 Because $\rho(n)\sim (\lambda n)^{-\beta}$ for large $n$, one sees immediately that $S_\infty^*$ is finite for $\beta>1$. For slow evanescence ($\lambda \to 0$) we find
 \be
 S_\infty^* \sim \frac{1-\mathcal{R}}{(\beta-1) \lambda}, \quad \beta>1.
 \ee
 For $\beta <1$ the result for slow evanescence is quite different.  For large $n$ we find
 \be
  S_n^* \sim \frac{1-\mathcal{R}}{1-\beta} \, \lambda^{-\beta}\,n^{1-\beta}, \quad 0<\beta<1.
\label{Sn3dpot}
\ee
For the marginal case $\beta=1$,
$S_n^* \sim (1-\mathcal{R}) \lambda^{-1 }\log n$.
For $\beta=0$ (no evanescence)
one recovers the classical result whereby $S_n$ is proportional to
$n$ \cite{Hughes1995}.
The average number of distinct sites visited by a mortal walker before it dies is thus finite for $\beta\ge 1$, whereas it
is infinite for $\beta<1$.  This is true for  $d$-dimensional lattices with $d\ge 2$.  However, for the one-dimensional lattice the  critical value is  $\beta=1/2$.

{\bf Mortal Brownian particles in continuous space, stretched exponential relaxation.}
It is well known that
$S_n$ for immortal walkers can be used to calculate the average volume $v_t$ of the Wiener sausage generated up to time $t$ by an immortal Brownian particle in a continuous medium. Since the relation between $S_n$ and $v_t$ is purely geometric, it can immediately be translated to
mortal walkers. Therefore, our results for $S_n^*$ can  be used to find the average volume $v_t^*$ of the Wiener sausage generated by a mortal Brownian particle up to time $t$. Explicitly, $S_n^*$ with $n\gg 1$ for a walker in a $d$-dimensional simple cubic lattice with lattice constant $\ell$, and  the Wiener sausage volume $v_t^*$ generated by a spherical diffusing particle of radius $R\gg \ell$ up to time $t=n\ell^2/(2dD)$ are related by  $v_t^*\sim \ell S_n^*$ for $d=1$  and by $v_t^*\sim \gamma_d (\ell/R)^2 R^d S_n^*$ for $d\ge 2$,  $\gamma_d$ being a constant that depends on dimension \cite{Berezhkovskii1989b}.

This connection
greatly expands the interesting world of stretched exponential relaxation discussed in the literature for several decades.
It is well-known that the evaluation of the number of distinct sites visited (or the volume explored) up to a given time is tantamount to the evaluation of the survival probability $\phi(t)$ up to that time of a fixed target particle of radius $R$ surrounded by a concentration of  diffusing point traps (target problem). The connection
is  $\phi(t)\sim \exp(-c v_t)$, which
also holds for evanescent traps with the replacement of
$v_t$ by $v_t^*$.
The identification of these traps as defects (i.e., carriers of free volume) is the basis of the defect diffusion model to explain the stretched exponential (or Kohlrausch-Williams-Watts) relaxation,
in which  $\ln \phi(t)\sim t^\theta$. However,  only the values $\theta=1/2$ and $\theta=1$ are possible for normal non-evanescent diffusive defects because  $S_n\propto v_t\propto t^{1/2}$ for $d=1$ (and then $\theta=1/2$) and $S_n\propto v_t\propto t$ for $d\ge 2$ (and then $\theta=1$).
This limited model \cite{Glarum1960} was extended
in \cite{Shlesinger1984} by assuming that the
movement of the defects can be described by a CTRW model with a power-law waiting time   $\psi(t)\sim t^{-1-\gamma}$, $0<\gamma<1$, which leads to $\theta=\gamma/2$ for $d=1$ and $\theta=\gamma$ for $d\neq 1$ \cite{Scher1991}.  That is, stretched exponential relaxation with $\theta\neq 1/2$ in this scenario is explained by assuming
anomalous diffusion of the defects, with
diffusion exponent $\gamma$ (leading to subdiffusion when $0<\gamma<1$).
In this context
we point to a recent
statistical model of random relaxation processes in disordered systems.   It provides a
general way to understand non-exponential relaxation processes \cite{Eliazar2012b,Eliazar2013}.

Our results
provide another route for explaining stretched exponential relaxation even for the case of normal defect diffusion by allowing the defects to disappear during the relaxation process \cite{Heggen2005,Chakravarty2009}. As we have shown, different kinds of evanescence lead to different laws of relaxation. For example, from Eq.~\eqref{Sn3dpot} we see that for $d\ge 3$, $v_t^*\propto t^{1-\beta}$ for $\beta<1$, so that one can arrive at stretched exponential relaxation with exponent $\theta=1-\beta$ when  the concentration of defects decays as a power law. Moreover, if the concentration of defects decays as $\rho(t)\sim 1/t$ for large $t$,
which corresponds to $\beta=1$, one finds
that $v_t^*\propto \ln t$, which in turns leads to \emph{algebraic} relaxation \cite{Shlesinger1979,Shlesinger1984a,BluKlaZuOptical}.

{\bf Trapping problem.}
The survival probability of the target in the target problem is frequently and appropriately used as a first approximation (the ``Rosenstock approximation") to the survival probability of the target in the so-called trapping problem in which the target diffuses and the traps are frozen \cite{Hughes1995}.
Our results
can also be applied to this  problem, now for the case of traps whose concentration decreases with time  \cite{Hollander1992}.

{\bf Conclusions.} As noted earlier,
it is surprising to find important solvable problems involving simple random walks, but we appear to have done so in the case of random walkers that evanesce in the course of their motion.  A number of classic problems such as the distinct number of sites visited as a function of time, or the survival probability of a target pursued by randomly walking traps, or any number of other quantities, change dramatically when the walkers can die in the course of their motion.  To mention but one or two such changes, we showed that the average number of distinct sites visited by an evanescing walker in $n$ steps as $n\to\infty$ may be finite (depending on the speed of evanescence), whereas it is clearly infinite if the walkers live forever.  Another, closely related to this, is the survival probability of a target in the presence of mobile traps.  If the traps live forever then the target will eventually disappear with certainty; if the traps evanesce, then the target may be spared.

 We have also enriched the world of stretched exponential relaxation of a target as calculated using the defect diffusion model. When the defects live forever, stretched exponential behavior in the CTRW model is only obtained if their diffusion is anomalous.  Here we have shown that the same stretched exponential behavior is obtained with normally diffusing defects provided they evanesce but do so sufficiently slowly.  Indeed, although not addressed here, it may happen that particles
are subdiffusive and \emph{also} have a finite lifetime
 \cite{Abad2010,Abad2012}.
Other problems for evanescent walkers related to explored territory (or distinct sites visited)
include the distribution of the distinct number of sites visited,
walks with different waiting time distributions (CTRWs), and L\'{e}vy flights and walks.
We are currently pursuing these and other related problems.

This work was partially supported by the Ministerio de Ciencia y
Tecnolog\'{\i}a (Spain) through Grant No. FIS2010-16587, by the Junta de
Extremadura (Spain) through Grant No. GRU10158, and by the
National Science Foundation under grant No. PHY-0855471.




%

\end{document}